\documentclass[fleqn,twoside]{article}
\usepackage[headings]{espcrc2}
\usepackage{amsmath,graphicx}
\title{Three-loop results in HQET}

\author{A.G.~Grozin\address{Institut f\"ur Theoretische Teilchenphysik,
Universit\"at Karlsruhe}}
       
\begin{document}

\begin{abstract}
Recent results and methods of three-loop calculations in HQET are reviewed.
\vspace{1pc}
\end{abstract}

\maketitle

\section{Off-shell HQET propagator diagrams}
\label{S:H3}

Two-loop HQET propagator diagrams were reduced~\cite{BG:91}
to two master integrals,
using integration by parts~\cite{CT:81} identities.

\begin{figure*}
\begin{center}
\begin{picture}(151,56.2)
\put(17,49.2){\makebox(0,0){\includegraphics{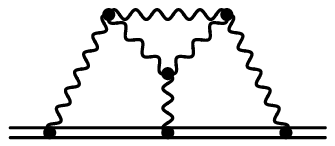}}}
\put(56,49.2){\makebox(0,0){\includegraphics{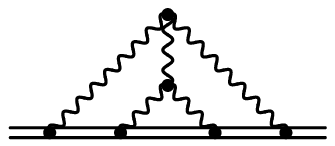}}}
\put(95,47.2){\makebox(0,0){\includegraphics{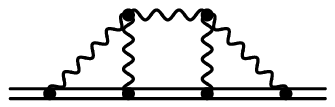}}}
\put(134,47.2){\makebox(0,0){\includegraphics{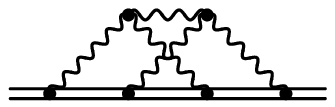}}}
\put(17,29.6){\makebox(0,0){\includegraphics{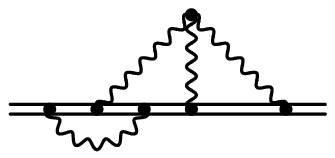}}}
\put(56,29.6){\makebox(0,0){\includegraphics{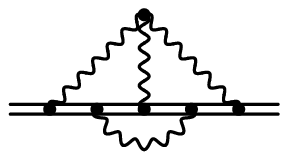}}}
\put(95,27.8){\makebox(0,0){\includegraphics{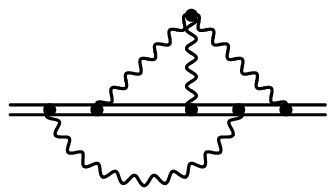}}}
\put(17,7.3){\makebox(0,0){\includegraphics{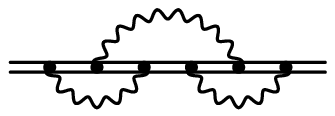}}}
\put(56,7.3){\makebox(0,0){\includegraphics{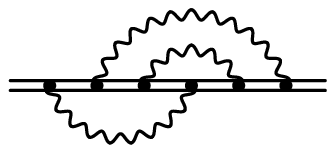}}}
\put(95,9.1){\makebox(0,0){\includegraphics{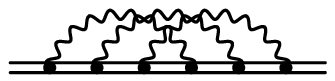}}}
\end{picture}
\end{center}
\caption{Topologies of three-loop HQET propagator diagrams}
\label{F:H3t}
\end{figure*}

\begin{figure*}
\begin{picture}(160,65.5)
\put(21,56.375){\makebox(0,0){\includegraphics{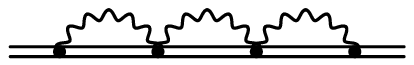}}}
\put(43,56.375){\makebox(0,0)[l]{{${}=I_1^3$}}}
\put(78,58.25){\makebox(0,0){\includegraphics{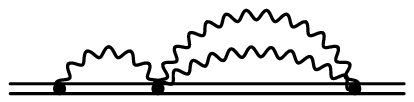}}}
\put(100,56.375){\makebox(0,0)[l]{{${}=I_1 I_2$}}}
\put(130,59.5){\makebox(0,0){\includegraphics{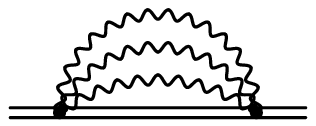}}}
\put(147,56.375){\makebox(0,0)[l]{{${}=I_3$}}}
\put(16,43.75){\makebox(0,0){\includegraphics{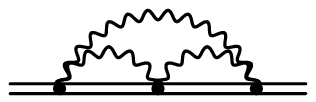}}}
\put(33,43.75){\makebox(0,0)[l]{{$\displaystyle{}\sim I_3\frac{I_1^2}{I_2}$}}}
\put(68,43.75){\makebox(0,0){\includegraphics{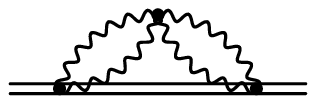}}}
\put(85,43.75){\makebox(0,0)[l]{{$\displaystyle{}\sim I_3\frac{G_1^2}{G_2}$}}}
\put(21,25.5){\makebox(0,0){\includegraphics{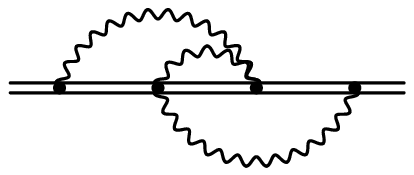}}}
\put(43,25.5){\makebox(0,0)[l]{{${}=I_1 J(1,1,-1+2\varepsilon,1,1)$}}}
\put(110,30.5){\makebox(0,0){\includegraphics{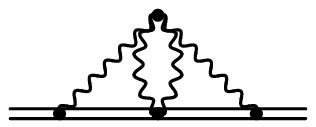}}}
\put(127,25.5){\makebox(0,0)[l]{{${}=G_1 I(1,1,1,1,\varepsilon)$}}}
\put(16,6){\makebox(0,0){\includegraphics{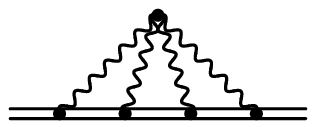}}}
\end{picture}
\caption{Master integrals for three-loop HQET propagator diagrams}
\label{F:H3m}
\end{figure*}

There are 10 generic topologies of three-loop
HQET propagator diagrams (Fig.~\ref{F:H3t}).
They can be reduced~\cite{G:00},
using integration by parts relations,
to 8 master integrals (Fig.~\ref{F:H3m}).
The algorithm has been constructed by hand,
and implemented as a REDUCE package Grinder~\cite{G:00}.
It is analogous to the massless package Mincer~\cite{Mincer}.

The first 5 master integrals can be easily expressed
via $\Gamma$ functions.
The next two ones are two-loop integrals
with a single $d$-dependent index.
The last one is truly three-loop.

\begin{figure}[h]
\begin{center}
\begin{picture}(75,5)
\put(21,0.5){\makebox(0,0){\includegraphics{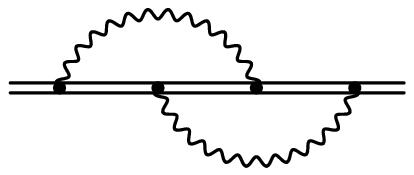}}}
\put(11,-2){\makebox(0,0){$n_1$}}
\put(31,3){\makebox(0,0){$n_2$}}
\put(21,3){\makebox(0,0){$n_3$}}
\put(16,10){\makebox(0,0){$n_4$}}
\put(26,-9){\makebox(0,0){$n_5$}}
\put(59,0.5){\makebox(0,0){\includegraphics{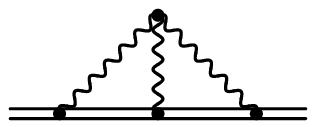}}}
\put(61,-0.5){\makebox(0,0){$n$}}
\end{picture}
\end{center}
\caption{Two-loop propagator diagrams}
\label{F:H3my}
\end{figure}

The first two-loop diagram (Fig.~\ref{F:H3my})
in coordinate space is just a double integral,
and can be easily calculated
for arbitrary $n_{1\ldots5}$~\cite{G:00}:
\begin{gather}
J(n_1,n_2,n_3,n_4,n_5) ={}
\nonumber\\
\Gamma(n_1+n_2+n_3+2(n_4+n_5-d))
\nonumber\\
\frac{\Gamma(n_1+n_3+2n_4-d)\Gamma(d/2-n_4)\Gamma(d/2-n_5)}%
{\Gamma(n_1+n_2+n_3+2n_4-d)\Gamma(n_4)\Gamma(n_5)\Gamma(n_1+n_3)}
\nonumber\\
{}_3F_2 \left(
\begin{array}{c}
n_1,d-2n_5,n_1+n_3+2n_4-d\\
n_1+n_3,n_1+n_2+n_3+2n_4-d
\end{array}
\right|\left.\vphantom{\frac{1}{1}}1\right)\,.
\raisetag{7mm}\label{H3:my}
\end{gather}

The second diagram in Fig.~\ref{F:H3my}
for arbitrary $n$ has been calculated~\cite{BB:94}
using Gegenbauer polynomials in $x$-space~\cite{CKT:80}:
\begin{gather}
I(1,1,1,1,n) =
\frac{\Gamma\left(\frac{d}{2}-1\right)\Gamma\left(\frac{d}{2}-n-1\right)}%
{\Gamma(d-2)}
\nonumber\\
\Biggl[ 2
\frac{\Gamma(2n-d+3)\Gamma(2n-2d+6)}{(n-d+3)\Gamma(3n-2d+6)}
\nonumber\\
{}\quad{}_3 F_2 \left(
\begin{array}{c}n-d+3,n-d+3,2n-2d+6\\n-d+4,3n-2d+6\end{array}
\right|\left.\vphantom{\frac{1}{1}}1\right)
\nonumber\\
{}\quad{} - \Gamma(d-n-2) \Gamma^2(n-d+3) \Biggr]\,.
\raisetag{11mm}\label{H3:bb}
\end{gather}

Similarly to the two-loop case~\cite{BG:95a},
some HQET propagator integrals are related
to on-shell massive propagator integrals by inversion
of Euclidean integration momenta:
\begin{gather}
\raisebox{-8.5mm}{\begin{picture}(42,23)
\put(21,11.5){\makebox(0,0){\includegraphics{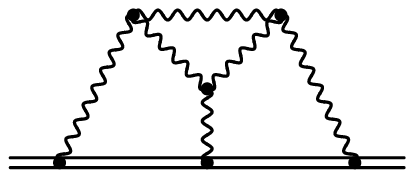}}}
\put(13.5,0){\makebox(0,0)[b]{$n_1$}}
\put(28.5,0){\makebox(0,0)[b]{$n_2$}}
\put(8.5,11.5){\makebox(0,0)[r]{$n_3$}}
\put(33,11.5){\makebox(0,0)[l]{$n_4$}}
\put(22,7.75){\makebox(0,0)[l]{$n_5$}}
\put(21,20){\makebox(0,0)[b]{$n_6$}}
\put(17.7,13){\makebox(0,0)[r]{$n_7$}}
\put(25,13){\makebox(0,0)[l]{$n_8$}}
\end{picture}} =
\nonumber\\
\raisebox{-8.5mm}{\begin{picture}(58,23)
\put(29,11.5){\makebox(0,0){\includegraphics{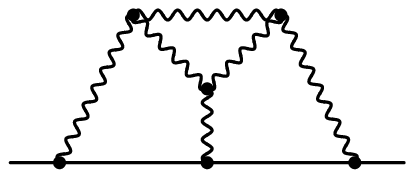}}}
\put(21.5,0){\makebox(0,0)[b]{$n_1$}}
\put(36.5,0){\makebox(0,0)[b]{$n_2$}}
\put(19,17){\makebox(0,0)[r]{$d-n_1-n_3$}}
\put(17,13){\makebox(0,0)[r]{${}-n_5-n_7$}}
\put(38.5,17){\makebox(0,0)[l]{$d-n_2-n_4$}}
\put(40,13){\makebox(0,0)[l]{${}-n_5-n_8$}}
\put(30,7.75){\makebox(0,0)[l]{$n_5$}}
\put(29,20){\makebox(0,0)[b]{$d-n_6-n_7-n_8$}}
\put(25.7,13){\makebox(0,0)[r]{$n_7$}}
\put(33,13){\makebox(0,0)[l]{$n_8$}}
\end{picture}}
\nonumber\\
\raisebox{-8.5mm}{\begin{picture}(42,18)
\put(21,9){\makebox(0,0){\includegraphics{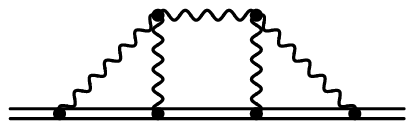}}}
\put(11,0){\makebox(0,0)[b]{$n_1$}}
\put(21,0){\makebox(0,0)[b]{$n_3$}}
\put(31,0){\makebox(0,0)[b]{$n_2$}}
\put(9.3,9){\makebox(0,0)[r]{$n_4$}}
\put(32.2,9){\makebox(0,0)[l]{$n_5$}}
\put(16.8,9){\makebox(0,0)[l]{$n_6$}}
\put(25.2,9){\makebox(0,0)[r]{$n_7$}}
\put(21,15){\makebox(0,0)[b]{$n_8$}}
\end{picture}} =
\nonumber\displaybreak\\
\raisebox{-8.5mm}{\begin{picture}(58,18)
\put(29,9){\makebox(0,0){\includegraphics{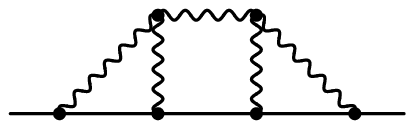}}}
\put(19,0){\makebox(0,0)[b]{$n_1$}}
\put(29,0){\makebox(0,0)[b]{$n_3$}}
\put(39,0){\makebox(0,0)[b]{$n_2$}}
\put(19.5,13){\makebox(0,0)[r]{$d-n_1-n_4$}}
\put(15.5,8.5){\makebox(0,0)[r]{${}-n_6$}}
\put(38,13){\makebox(0,0)[l]{$d-n_2-n_5$}}
\put(41,8.5){\makebox(0,0)[l]{${}-n_7$}}
\put(24.8,9){\makebox(0,0)[l]{$n_6$}}
\put(33.2,9){\makebox(0,0)[r]{$n_7$}}
\put(29,15){\makebox(0,0)[b]{$d-n_3-n_6-n_7-n_8$}}
\end{picture}}
\label{H3:inv}
\end{gather}
In particular, at $d=4$ we have
\begin{gather}
\raisebox{-4.5mm}{\includegraphics{h3t3.eps}} =
\raisebox{-4.5mm}{\includegraphics{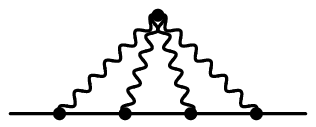}}
\nonumber\\
{} = - 5 \zeta_5 + 12 \zeta_2 \zeta_3\,.
\label{H3:inv4}
\end{gather}
Reducing the left-hand side to the master integrals,
one can express~\cite{CM:02} the last master integral
in Fig.~\ref{F:H3m} via the right-hand side of~(\ref{H3:inv4}),
which is an on-shell master integral known from~\cite{LR:96}.
Note that $\mathcal{O}(\varepsilon)$ corrections
to~(\ref{H3:inv4}) are not known.

Using this technique, the HQET heavy-quark propagator
has been calculated up to three loops~\cite{CG:03},
and the heavy-quark field anomalous dimension
(obtained earlier by a completely different method~\cite{MR:00})
has been confirmed.
The anomalous dimension of the HQET heavy--light quark current
has been calculated~\cite{CG:03}.
The correlator of two heavy--light currents has been found,
up to three loops, including light-quark mass corrections
of order $m$ and $m^2$~\cite{CG2}.
The quark-condensate contribution to this correlator
has been also calculated up to three loops~\cite{CG2}.
Its ultraviolet divergence yields the difference
of twice the anomalous dimension of the heavy-quark current
and the that of the quark condensate,
thus providing a completely independent confirmation
of the result obtained in~\cite{CG:03}.
The gluon-condensate contribution has been calculated
up to two loops~\cite{CG2}
(at one loop it vanishes).

\section{On-shell HQET propagator diagrams with mass}
\label{S:M3}

\subsection{Reduction}
\label{S:M3r}

The two-loop on-shell HQET propagator integral
with a massive loop can be expressed via $\Gamma$ functions
for arbitrary indices~\cite{GSS:06}:
\begin{gather}
\raisebox{-7mm}{\begin{picture}(27,15)
\put(13.5,9){\makebox(0,0){\includegraphics{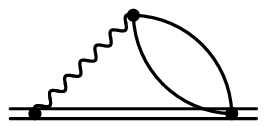}}}
\put(13.5,0){\makebox(0,0)[b]{$n_1$}}
\put(6,11){\makebox(0,0){$n_2$}}
\put(13.5,7){\makebox(0,0){$n_3$}}
\put(24,11){\makebox(0,0){$n_4$}}
\end{picture}} =
\nonumber\displaybreak\\
\frac{\Gamma\left(\frac{n_1}{2}\right)
\Gamma\left(\frac{n_1-d}{2}+n_2+n_3\right)
\Gamma\left(\frac{n_1-d}{2}+n_2+n_4\right)}%
{2\Gamma(n_1)\Gamma(n_3)\Gamma(n_4)
\Gamma(n_1+2n_2+n_3+n_4-d)}
\nonumber\\
\frac{\Gamma\Bigl(\frac{n_1}{2}+n_2+n_3+n_4-d\Bigr)
\Gamma\left(\frac{d-n_1}{2}-n_2\right)}%
{\Gamma\left(\frac{d-n_1}{2}\right)}\,.
\label{M3:l2}
\end{gather}

\begin{figure}[h]
\begin{center}
\begin{picture}(61,1)
\put(14,0){\makebox(0,0){\includegraphics{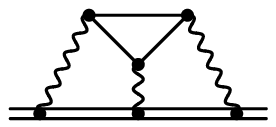}}}
\put(47,0){\makebox(0,0){\includegraphics{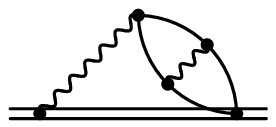}}}
\end{picture}
\end{center}
\caption{Topologies of on-shell HQET propagator diagrams with mass}
\label{F:M3t}
\end{figure}

There are two generic topologies of three-loop
on-shell HQET propagator diagrams with a massive loop
(Fig.~\ref{F:M3t}).
Algorithms of their reduction to master integrals,
using integration by parts identities,
have been constructed~\cite{GSS:06}
by Gr\"obner bases technique~\cite{SS:06}.
All scalar integrals can be divided into two classes
which are not mixed by recurrence relations:
apparently even and apparently odd ones.
They are integrals which would be even or odd
with respect to $v\to-v$
if there were no $i0$ in denominators.

All apparently even integrals of the first topology
reduce to
\begin{equation*}
\begin{split}
&\raisebox{-6.3mm}{\includegraphics{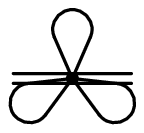}}\quad
\raisebox{-6.3mm}{\includegraphics{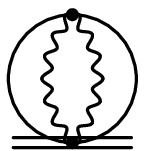}}\quad
\raisebox{-4.3mm}{\includegraphics{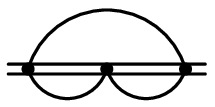}}\\
&\raisebox{-3.8mm}{\includegraphics{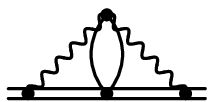}}\quad
\raisebox{-3.8mm}{\includegraphics{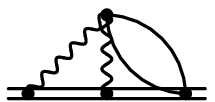}}
\end{split}
\end{equation*}
while apparently odd ones to
\begin{equation*}
\raisebox{-6.3mm}{\includegraphics{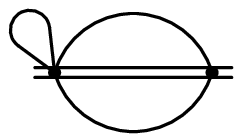}}\quad
\raisebox{-5.8mm}{\includegraphics{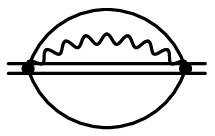}}\quad
\raisebox{-3.8mm}{\includegraphics{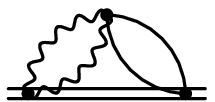}}
\end{equation*}
All apparently even integrals of the second topology
reduce to
\begin{equation*}
\raisebox{-6.3mm}{\includegraphics{i03.eps}}\quad
\raisebox{-6.3mm}{\includegraphics{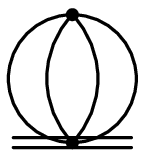}}
\end{equation*}
while apparently odd ones to
\begin{equation*}
\raisebox{-6.3mm}{\includegraphics{ij0.eps}}\quad
\raisebox{-5.8mm}{\includegraphics{j1.eps}}\quad
\raisebox{-3.8mm}{\includegraphics{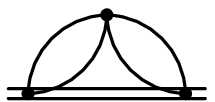}}
\end{equation*}

The master integrals
\begin{equation*}
\raisebox{-6.3mm}{\includegraphics{i1.eps}}\quad
\raisebox{-5.8mm}{\includegraphics{j1.eps}}\quad
\raisebox{-5.8mm}{\includegraphics{j2.eps}}
\end{equation*}
can be easily expressed via $\Gamma$ functions.
The master integral
\begin{equation*}
\raisebox{-6.3mm}{\includegraphics{i5.eps}}
\end{equation*}
has been investigated in detail~\cite{B:92,B:96}.

\subsection{A master integral}
\label{S:M31}

Now we shall discuss the integrals
\begin{equation}
I_{n_1 n_2 n_3} =
\raisebox{-10mm}{\begin{picture}(27,17.25)
\put(13.5,8.625){\makebox(0,0){\includegraphics{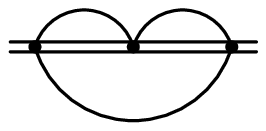}}}
\put(13.5,0){\makebox(0,0)[b]{$n_3$}}
\put(8.5,17.25){\makebox(0,0)[t]{$n_2$}}
\put(18.5,17.25){\makebox(0,0)[t]{$n_2$}}
\put(8.5,9.5){\makebox(0,0)[t]{$n_1$}}
\put(18.5,9.5){\makebox(0,0)[t]{$n_1$}}
\end{picture}}
\label{M3:In1n2n3}
\end{equation}
($I_{111}$ is one of the master integrals).
Several approaches have been tried~\cite{GSS:06,GHM:07}.
The best result was obtained~\cite{GHM:07}
using a method similar to~\cite{B:92}.

First we consider the one-loop subdiagram
\begin{gather}
I_{n_1 n_2}(p_0) =
\raisebox{-2.5mm}{\begin{picture}(30,14)
\put(15,7){\makebox(0,0){\includegraphics{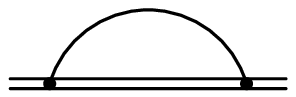}}}
\put(15,0){\makebox(0,0)[b]{$n_1$}}
\put(15,14){\makebox(0,0)[t]{$n_2$}}
\end{picture}} =
\label{M3:In1n2}\\
\frac{1}{i\pi^{d/2}} \int
\frac{d k_0\,d^{d-1}\vec{k}}{[-2(k_0+p_0)-i0]^{n_1} [1-k^2-i0]^{n_2}}\,.
\nonumber
\end{gather}
After the Wick rotation, we integrate in $d^{d-1}\vec{k}$:
\begin{gather*}
I_{n_1 n_2}(p_0) =
\frac{\Gamma(n_2-(d-1)/2)}{\pi^{1/2} \Gamma(n_2)}\times\\
{}\qquad\int_{-\infty}^{+\infty} d k_{E0}
\frac{(k_{E0}^2+1)^{(d-1)/2-n_2}}{(- 2 p_0 - 2 i k_{E0})^{n_1}}\,.
\end{gather*}
If $p_0<0$, we can deform the integration contour
\begin{gather*}
\begin{picture}(42,42)
\put(21,21){\makebox(0,0){\includegraphics{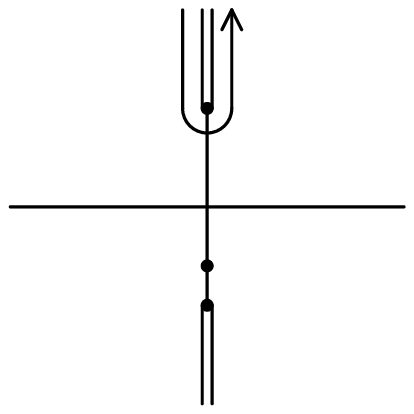}}}
\put(22,15){\makebox(0,0)[l]{$i p_0$}}
\put(18,31){\makebox(0,0)[r]{$i$}}
\put(18,11){\makebox(0,0)[r]{$-i$}}
\put(25,39){\makebox(0,0)[l]{C}}
\put(35,31){\makebox(0,0){$k_{E0}$}}
\end{picture}\\
I_{n_1 n_2}(p_0) = 2 \frac{\Gamma(n_2-(d-1)/2)}{\pi^{1/2} \Gamma(n_2)}
\times\\
\cos\left[\pi \left(\frac{d}{2} - n_2\right)\right]
\int_1^\infty dk
\frac{(k^2-1)^{(d-1)/2-n_2}}{(2k - 2p_0)^{n_1}}\,.
\end{gather*}
This integral is
\begin{gather*}
I_{n_1 n_2}(p_0) =\\
\frac{\Gamma(n_1+n_2-2+\varepsilon) \Gamma(n_1+2n_2-4+2\varepsilon)}%
{\Gamma(n_2) \Gamma(2(n_1+n_2-2+\varepsilon))}\times\\
{}_2 F_1 \left( \left.
\begin{array}{c}
n_1,n_1+2n_2-4+2\varepsilon\\
n_1+n_2-\frac{3}{2}+\varepsilon
\end{array}
\right| \frac{1}{2} \left(1 + p_0\right) \right)\,,
\end{gather*}
or, after using a ${}_2 F_1$ identity,
\begin{gather}
I_{n_1 n_2}(p_0) =
\nonumber\\
\frac{\Gamma(n_1+n_2-2+\varepsilon) \Gamma(n_1+2n_2-4+2\varepsilon)}%
{\Gamma(n_2) \Gamma(2(n_1+n_2-2+\varepsilon))}\times
\nonumber\\
{}_2 F_1 \left( \left.
\begin{array}{c}
\frac{1}{2} n_1, \frac{1}{2} n_1 + n_2 - 2 + \varepsilon\\
n_1 + n_2 - \frac{3}{2} + \varepsilon
\end{array}
\right| 1 - p_0^2 \right)\,.
\label{M3:res1}
\end{gather}
This result was obtained~\cite{GHM:07}
using the HQET Feynman parametrization.

Now we can integrate in $d^{d-1}\vec{p}$
in the three-loop diagram:
\begin{gather}
I_{n_1 n_2 n_3} =
\frac{\Gamma(n_3-3/2+\varepsilon)}{\pi^{1/2}\Gamma(n_3)}\times
\nonumber\\
\int_{-\infty}^{+\infty}
I_{n_1 n_2}^2(i p_{E0}) (1+p_{E0}^2)^{3/2-n_3-\varepsilon} d p_{E0}\,.
\label{M3:I}
\end{gather}
The square of ${}_2 F_1$ in~(\ref{M3:res1})
can be expressed via an ${}_3 F_2$
using the Clausen identity.
We analytically continue this ${}_3 F_2$ from $1+p_{E0}^2>1$
to $z=1/(1+p_{E0}^2)<1$ and integrate~(\ref{M3:I}) term by term.
The result contains, in general, three ${}_4 F_3$
of unit argument.

The convergent integral $I_{122}$ is related
to the master integral $I_{111}$ by
\begin{equation*}
I_{122} = -
\frac{(d-3)^2 (d-4) (3d-8) (3d-10)}{8 (3d-11) (3d-13)} I_{111}
\end{equation*}
For this integral, we obtain~\cite{GHM:07}
\begin{gather}
\frac{I_{122}}{\Gamma^3(1+\varepsilon)} =
- \frac{1}{2\varepsilon^2} \Biggl[ \frac{1}{1+2\varepsilon}
\nonumber\\
{}\qquad{}_4 F_3 \left( \left.
\begin{array}{c}
1, \frac{1}{2}-\varepsilon, 1+\varepsilon, -2\varepsilon\\
\frac{3}{2}+\varepsilon, 1-\varepsilon, 1-2\varepsilon
\end{array}
\right| 1 \right)
\nonumber\\
{} - \frac{2}{1+4\varepsilon}
\frac{\Gamma^2(1-\varepsilon) \Gamma^3(1+2\varepsilon)}%
{\Gamma^2(1+\varepsilon) \Gamma(1-2\varepsilon) \Gamma(1+4\varepsilon)}
\nonumber\\
{}\qquad{}_3 F_2 \left( \left.
\begin{array}{c}
\frac{1}{2}, 1+2\varepsilon, -\varepsilon\\
\frac{3}{2}+2\varepsilon, 1-\varepsilon
\end{array}
\right| 1 \right)
\label{M3:hyper}\\
{} + \frac{1}{1+6\varepsilon}
\nonumber\\
{}\qquad
\frac{\Gamma^2(1-\varepsilon) \Gamma^4(1+2\varepsilon)
\Gamma(1-2\varepsilon) \Gamma^2(1+3\varepsilon)}%
{\Gamma^4(1+\varepsilon) \Gamma(1+4\varepsilon)
\Gamma(1-4\varepsilon) \Gamma(1+6\varepsilon)}
\Biggr]\,.
\nonumber
\end{gather}
Expansion of this result up to $\varepsilon^7$
agrees with~\cite{GHM:07}
\begin{equation}
\frac{I_{122}}{\Gamma^3(1+\varepsilon)} =
\frac{\pi^2}{3}
\frac{\Gamma^3(1+2\varepsilon) \Gamma^2(1+3\varepsilon)}%
{\Gamma^6(1+\varepsilon) \Gamma(2+6\varepsilon)}\,.
\label{M3:Maitre}
\end{equation}
This equality has also been checked by high precision
numerical calculations at some finite $\varepsilon$ values.
During the workshop, David Broadhurst has rewritten this conjectured
hypergeometric identity in a nice form
\begin{gather}
b(\varepsilon) =
\frac{\Gamma(1-\varepsilon)\Gamma(1+2\varepsilon)}{\Gamma(1+\varepsilon)}\,,
\nonumber\\
g_n(\varepsilon) =
\frac{b^n(\varepsilon)}{b(n\varepsilon)(1+2n\varepsilon)}\,,
\nonumber\\
g_1(\varepsilon)
{}_4 F_3 \left( \left.
\begin{array}{c}
1, \frac{1}{2}-\varepsilon, 1+\varepsilon, -2\varepsilon\\
\frac{3}{2}+\varepsilon, 1-\varepsilon, 1-2\varepsilon
\end{array}
\right| 1 \right)
\nonumber\\
{}\qquad{} - 2 g_2(\varepsilon)
{}_3 F_2 \left( \left.
\begin{array}{c}
\frac{1}{2}, 1+2\varepsilon, -\varepsilon\\
\frac{3}{2}+2\varepsilon, 1-\varepsilon
\end{array}
\right| 1 \right)
\nonumber\\
{}\qquad{} + g_3(\varepsilon) = 0\,.
\label{M3:David}
\end{gather}
We have no analytical proof.

\subsection{Other master integrals}
\label{S:M32}

Now we turn to other master integrals~\cite{GSS:06}.
By applying Mellin--Barnes techniques
to the $\alpha$-representation
and using the Barnes lemmas,
we succeeded in calculating one integral exactly:
\begin{gather}
\raisebox{-3.8mm}{\includegraphics{i3.eps}} =
\nonumber\\
\frac{\Gamma(1/2-\varepsilon)
\Gamma(-\varepsilon)
\Gamma^2(2\varepsilon)
\Gamma(1+\varepsilon)
\Gamma(3\varepsilon-1)}%
{4 \Gamma(3/2-\varepsilon) \Gamma(4\varepsilon)}
\nonumber\\
\left[ \psi(1/2-\varepsilon) + \psi(1-\varepsilon) - 2 \log 2
+2 \gamma_{\text{E}} \right]\,.
\label{M3:i3}
\end{gather}

\begin{figure*}
\begin{center}
\begin{picture}(151,56.2)
\put(17,49.2){\makebox(0,0){\includegraphics{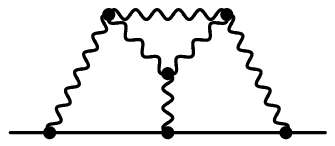}}}
\put(56,49.2){\makebox(0,0){\includegraphics{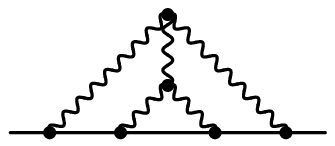}}}
\put(95,47.2){\makebox(0,0){\includegraphics{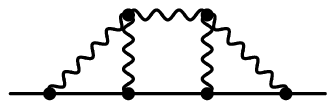}}}
\put(134,47.2){\makebox(0,0){\includegraphics{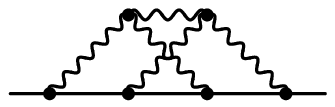}}}
\put(17,29.6){\makebox(0,0){\includegraphics{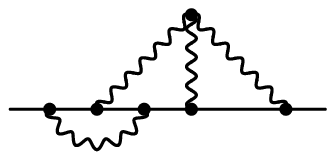}}}
\put(56,29.6){\makebox(0,0){\includegraphics{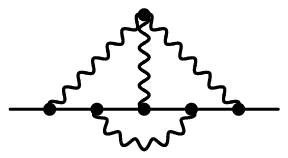}}}
\put(95,27.8){\makebox(0,0){\includegraphics{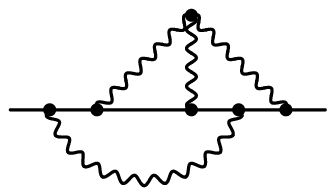}}}
\put(17,7.3){\makebox(0,0){\includegraphics{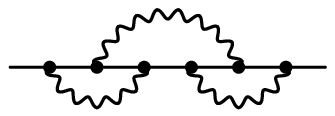}}}
\put(56,7.3){\makebox(0,0){\includegraphics{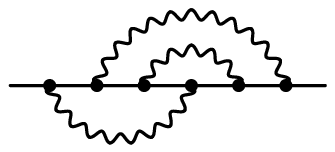}}}
\put(95,9.1){\makebox(0,0){\includegraphics{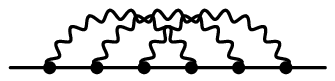}}}
\put(134,12.4){\makebox(0,0){\includegraphics{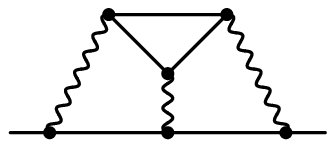}}}
\end{picture}
\end{center}
\caption{Topologies of three-loop massive on-shell propagator diagrams}
\label{F:O3t}
\end{figure*}

In other master integral calculations,
we use the Mellin--Barnes of the massive two-point function:
\begin{gather}
\raisebox{-8mm}{\begin{picture}(22,18)
\put(11,9){\makebox(0,0){\includegraphics{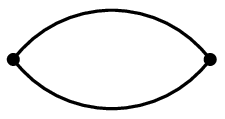}}}
\put(11,16){\makebox(0,0){$n_1$}}
\put(11,2){\makebox(0,0){$n_2$}}
\end{picture}} =
\nonumber\\
\frac{1}{i\pi^{d/2}}
\int \frac{d^d k}{\left[m^2-k^2-i0\right]^{n_1}
\left[m^2-(k+p)^2-i0\right]^{n_2}}
\nonumber\\
{} = \frac{m^{d-2(n_1+n_2)}}{\Gamma(n_1) \Gamma(n_2)}
\frac{1}{2\pi i} \int_{-i\infty}^{+i\infty} dz\,\Gamma(-z)\times
\nonumber\\
{}\qquad\frac{\Gamma(n_1+z) \Gamma(n_2+z)
\Gamma(n_1+n_2-d/2+z)}{\Gamma(n_1+n_2+2z)}\times
\nonumber\\
{}\qquad m^{-2z}
\raisebox{-3mm}{\begin{picture}(22,8)
\put(11,4){\makebox(0,0){\includegraphics{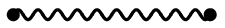}}}
\put(11,6){\makebox(0,0){$-z$}}
\end{picture}}
\label{M3:MB}
\end{gather}

Applying this representation
and eliminating one of the lines
in the triangle with unit indices
using integration by parts,
we obtain a single Mellin--Barnes integral:
\begin{gather}
\raisebox{-3.8mm}{\includegraphics{i4.eps}} =
\frac{\Gamma^2(2\varepsilon)\Gamma(3\varepsilon-1)}{4\Gamma(4\varepsilon)}
\times
\nonumber\\
\frac{1}{2\pi i} \int_{-i\infty}^{+i\infty} dz\,
\Gamma(-z) \Gamma(-2\varepsilon-z) \Gamma(-\varepsilon-z)\times
\nonumber\\
{}\qquad
\frac{\Gamma(1+z) \Gamma(1/2+\varepsilon+z) \Gamma(1+\varepsilon+z)}%
{\Gamma(1-2\varepsilon-z) \Gamma(3/2+\varepsilon+z)} =
\nonumber\\
- \Gamma^3(1+\varepsilon) \Biggl[
\frac{\pi^2}{9\varepsilon^2}
- \frac{6\zeta_3-5\pi^2}{9\varepsilon}
\nonumber\\
{}\qquad{} + \frac{11}{270} \pi^4 - \frac{10}{3} \zeta_3 + \frac{19}{9} \pi^2
\nonumber\displaybreak\\
{} + \left( - \frac{8}{3} \zeta_5 + \frac{8}{9} \pi^2 \zeta_3
+ \frac{11}{54} \pi^4 - \frac{38}{3} \zeta_3 + \frac{65}{9} \pi^2
\right) \varepsilon
\nonumber\\
{} + \cdots \Biggr]\,.
\label{M3:i4}
\end{gather}

Similarly, in this case,
we apply~(\ref{M3:MB}) once
and use~(\ref{M3:l2}):
\begin{gather}
\raisebox{-3.8mm}{\includegraphics{j3.eps}} =
\frac{\pi^{3/2} }{4^\varepsilon \Gamma(3/2-\varepsilon)}\times
\nonumber\\
\frac{1}{2\pi i} \int\limits_{-i\infty}^{+i\infty} dz
\frac{\Gamma(1+z) \Gamma(3/2-\varepsilon+z) \Gamma(\varepsilon+z)}%
{\Gamma(3/2+z) \Gamma(\varepsilon-z)}\times
\nonumber\\
{}\qquad
\Gamma(-z) \Gamma(-1/2+\varepsilon-z) \Gamma(-3/2+2\varepsilon-z)
\nonumber\\
{} = -\frac{32}{3} \pi^2 + \cdots
\label{M3:j3}
\end{gather}

Feynman integrals considered here were used~\cite{GSS:06}
for calculating the matching coefficients
for the HQET heavy-quark field and the heavy--light quark current
between the $b$-quark HQET with dynamic $c$-quark loops
and without such loops
(the later theory is the low-energy approximation
for the former one at scales below $m_c$).
Another recent application --- the effect of $m_c\ne0$
on $b\to c$ plus lepton pair at three loops~\cite{CP:08}.
The method of regions was used;
the purely soft region (loop momenta $\sim m_c$)
gives integrals of this type.
Two extra terms of $\varepsilon$ expansion
of the master integral of Sect.~\ref{S:M31}
were required for this calculation
which were not obtained in~\cite{GSS:06}.
This was the initial motivation for~\cite{GHM:07}.

\section{On-shell massive QCD propagator diagrams}
\label{S:O3}

These diagrams are used for calculation
of QCD/HQET matching coefficients~\cite{BG:95}.
There are two generic topologies
of two-loop on-shell propagator diagrams.
They were reduced~\cite{B:90,B:92,FT:92}
to three master integrals, using integration by parts.

\begin{figure*}
\begin{center}
\begin{picture}(158.4,147.2)
\put(11.2,132){\makebox(0,0){\includegraphics[scale=0.8]{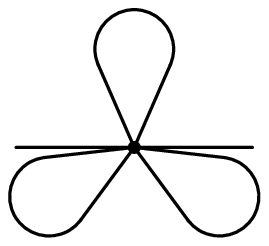}}}
\put(47.2,132){\makebox(0,0){\includegraphics[scale=0.8]{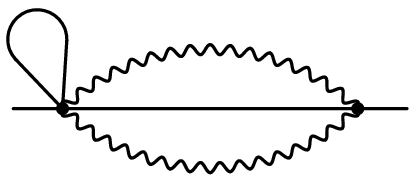}}}
\put(88.8,132){\makebox(0,0){\includegraphics[scale=0.8]{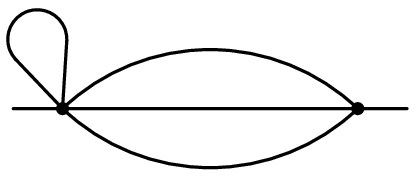}}}
\put(8.8,104){\makebox(0,0){\includegraphics[scale=0.8]{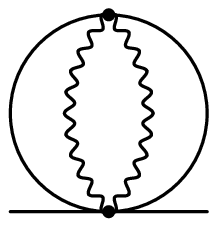}}}
\put(34.4,104){\makebox(0,0){\includegraphics[scale=0.8]{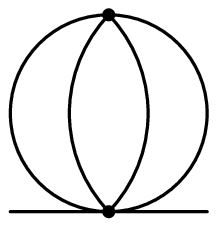}}}
\put(68,98.4){\makebox(0,0){\includegraphics[scale=0.8]{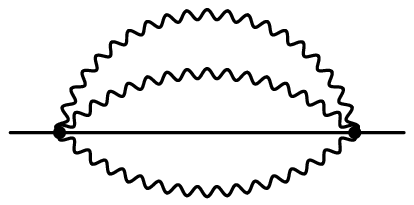}}}
\put(109.6,98.4){\makebox(0,0){\includegraphics[scale=0.8]{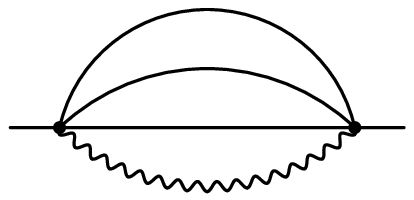}}}
\put(16.8,75.6){\makebox(0,0){\includegraphics[scale=0.8]{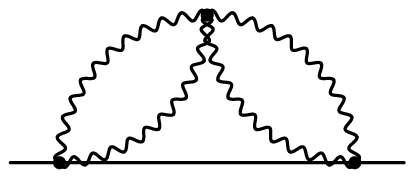}}}
\put(58.4,75.6){\makebox(0,0){\includegraphics[scale=0.8]{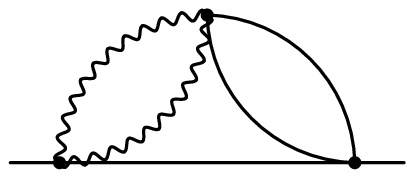}}}
\put(100,75.6){\makebox(0,0){\includegraphics[scale=0.8]{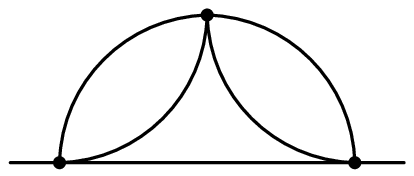}}}
\put(141.6,69){\makebox(0,0){\includegraphics[scale=0.8]{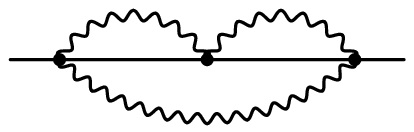}}}
\put(16.8,53.7){\makebox(0,0){\includegraphics[scale=0.8]{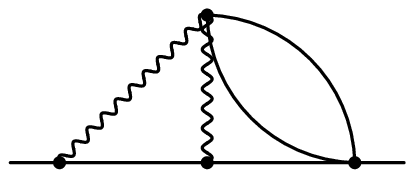}}}
\put(58.4,53.7){\makebox(0,0){\includegraphics[scale=0.8]{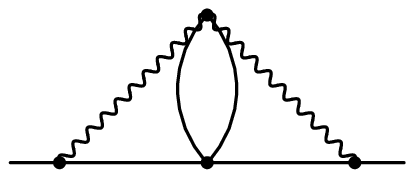}}}
\put(100,48.825){\makebox(0,0){\includegraphics[scale=0.8]{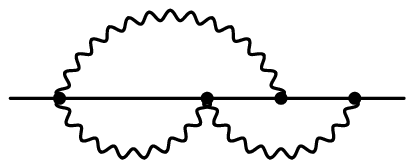}}}
\put(16.8,24.4){\makebox(0,0){\includegraphics[scale=0.8]{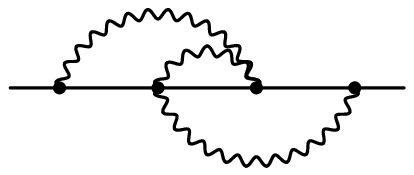}}}
\put(58.4,25.9){\makebox(0,0){\includegraphics[scale=0.8]{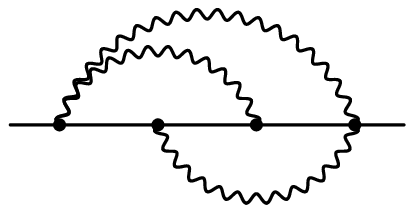}}}
\put(100,30.4){\makebox(0,0){\includegraphics[scale=0.8]{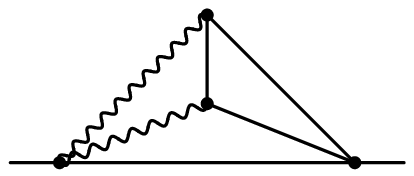}}}
\put(12.8,4.8){\makebox(0,0){\includegraphics[scale=0.8]{m7.eps}}}
\put(47.2,2.96){\makebox(0,0){\includegraphics[scale=0.8]{m3t10.eps}}}
\end{picture}
\end{center}
\caption{Master integrals for three-loop massive on-shell propagator diagrams}
\label{F:O3m}
\end{figure*}

There are 11 generic topologies of three-loop
HQET propagator diagrams (Fig.~\ref{F:O3t})
(10 of them are the same as in HQET, Fig.~\ref{F:H3t},
plus the diagram with a closed heavy-quark loop
which does not exist in HQET).
They can be reduced~\cite{MR:00} to 19 master integrals (Fig.~\ref{F:O3m})
using integration by parts.
The reduction algorithm was first implemented
as a FORM package SHELL3~\cite{MR:00};
in some papers~\cite{MMPS:07}, reduction is performed
using the Laporta method.
The master integrals are known from~\cite{LR:96,MR:00}
(see also~\cite{MMPS:07}).

As an example, let's consider chromomagnetic interaction
of the heavy quark in HQET.
The two-loop anomalous dimension
of the chromomagnetic interaction operator
has been calculated in~\cite{ABN:97},
using off-shell HQET diagrams (Sect.~\ref{S:H3})
and $R^*$ operation to get rid of infrared divergences.
It was confirmed~\cite{CG:97}, a week later,
using a completely different approach ---
QCD/HQET matching.
In the same paper,
also the coefficient of the chromomagnetic interaction operator
in the HQET Lagrangian has been calculated up to two loops.
Recently, both the anomalous dimension
and the interaction coefficient have been calculated
at three loops~\cite{GMPS:08},
also using the QCD/HQET matching.
These results contain two colour factors
which don't reduce to $C_F$ and $C_A$
(due to gluonic ``light-by-light'' diagrams;
in the $\beta$-function and basic anomalous dimensions of QCD,
these colour factors first appear at four loops.

I am grateful to
S.~Bekavac,
D.J.~Broadhurst,
K.G.~Chetyrkin,
A.~Czarnecki,
A.I.~Davydychev,
T.~Huber,
D.~Ma\^{\i}tre,
P.~Marquard,
J.H.~Piclum,
D.~Seidel,
A.V.~Smirnov,
V.A.~Smirnov,
M.~Steinhauser
for collaboration on multi-loop projects in HQET;
to the organizers of Loops and Legs 2008;
and to D.J.~Broadhurst for discussions
of the master integral of Sect.~\ref{S:M31}.

\end{document}